# Green Algorithms: Quantifying the carbon footprint of computation

Loïc Lannelongue^1,4,7, Jason Grealey^2,3, Michael Inouye*1,2,4,5,6,7,8

[1]Cambridge Baker Systems Genomics Initiative, Department of Public Health and Primary Care, University of Cambridge, Cambridge, UK
[2]Cambridge Baker Systems Genomics Initiative, Baker Heart and Diabetes Institute, Melbourne, Victoria, Australia
[3]Department of Mathematics and Statistics, La Trobe University, Melbourne, Australia
[4]British Heart Foundation Cardiovascular Epidemiology Unit, Department of Public Health and Primary Care, University of Cambridge, Cambridge, UK
[5]British Heart Foundation Centre of Research Excellence, University of Cambridge, Cambridge, UK
[6]National Institute for Health Research Cambridge Biomedical Research Centre, University of Cambridge and Cambridge University Hospitals, Cambridge, UK
[7]Health Data Research UK Cambridge, Wellcome Genome Campus and University of Cambridge, Cambridge, UK
[8]The Alan Turing Institute, London, UK

^ Joint first authors
* Correspondence: MI (mi336@medschl.cam.ac.uk; minouye@baker.edu.au)

## Abstract

Climate change is profoundly affecting nearly all aspects of life on earth, including human societies, economies and health. Various human activities are responsible for significant greenhouse gas emissions, including data centres and other sources of large-scale computation. Although many important scientific milestones have been achieved thanks to the development of high-performance computing, the resultant environmental impact has been underappreciated. In this paper, we present a methodological framework to estimate the carbon footprint of any computational task in a standardised and reliable way, based on the processing time, type of computing cores, memory available and the efficiency and location of the computing facility. Metrics to interpret and contextualise greenhouse gas emissions are defined, including the equivalent distance travelled by car or plane as well as the number of tree-months necessary for carbon sequestration. We develop a freely available online tool, Green Algorithms (www.green-algorithms.org), which enables a user to estimate and report the carbon footprint of their computation. The Green Algorithms tool easily integrates with computational processes as it requires minimal information and does not interfere with existing code, while also accounting for a broad range of CPUs, GPUs, cloud computing, local servers and desktop computers. Finally, by applying Green Algorithms, we quantify the greenhouse gas emissions of algorithms used for particle physics simulations, weather forecasts and natural language processing. Taken together, this study develops a simple generalisable framework and freely available tool to quantify the carbon footprint of nearly any computation. Combined with a series of recommendations to minimise unnecessary $CO_2$ emissions, we hope to raise awareness and facilitate greener computation.

# Introduction

The concentration of greenhouse gases in the atmosphere has a dramatic influence on climate change with both global and locally focused consequences, such as rising sea levels, devastating wildfires in Australia, extreme typhoons in the Pacific, severe droughts across Africa, as well as repercussions for human health.

With 100 megatonnes of $CO_2$ emissions per year[a], similar to American commercial aviation, the contribution of data centres and high-performance computing facilities to climate change is substantial. So far, rapidly increasing demand has been paralleled by increasingly energy-efficient facilities, with overall electricity consumption of data centres somewhat stable. However, this stability is likely to end in the coming years, with a best-case scenario forecasting a three-fold increase in the energy needs of the sector[1,2].

Advances in computation, including those in hardware, software and algorithms, have enabled scientific research to progress at unprecedented rates. Weather forecasts have increased in accuracy to the point where 5-day forecasts are approximately as accurate as 1-day forecasts 40 years ago[3], physics algorithms have produced the first direct image of a black hole 55 million light-years away[4–6], the human genome has been mined to uncover thousands of genetic variants for disease[7], and machine learning permeates many aspects of society, including economic and social interactions[8–11]. However, the costs associated with large-scale computation are not being fully captured.

Power consumption results in greenhouse gas (GHG) emissions and the environmental costs of performing computations using data centres, personal computers, and the immense diversity of architectures are unclear. While programmes in green computing (the study of environmentally responsible information and communications technologies) have been developed over the past decade, these mainly focus on energy-efficient hardware and cloud-related technologies[12–14].

With widely recognised power-hungry and expensive training algorithms, deep learning has begun to address its carbon footprint. Machine learning (ML) models have grown exponentially in size over the past few years[15], with some algorithms training for thousands of core-hours, and the associated energy consumption and cost have become a growing concern[16]. In natural language processing (NLP), Strubell et al.[17] found that designing and training translation engines can emit between 0.6 and 280 tonnes of $CO_2$. While not all NLP algorithms require frequent retraining, algorithms in other fields are run daily or weekly, multiplying their energy consumption.

Previous studies have made advances in estimating GHG emissions of computation but have limitations which preclude broad applicability. These limitations include the requirement that users self-monitor their power consumption[17] and are restricted with respect to hardware (e.g. GPUs and/or cloud systems[18,19]), software (e.g. Python package integration[19]), or applications (e.g. machine learning)[17–19]. To facilitate green computing and widespread user uptake, there is a clear, and arguably urgent, need for both a general and easy-to-use methodology for estimating carbon emissions that can be applied to any computational task.

In this study, we present a simple and widely applicable method and a tool for estimating the carbon footprint of computation. The method considers the different sources of energy usage, such as processors and memory, overhead of computing facilities and geographic location, while balancing accuracy and practicality. The online calculator (www.green-algorithms.org) implements this methodology and provides further context by interpreting carbon amounts using travel distances and carbon sequestration. We demonstrate the applicability of the Green Algorithms method by estimating the carbon footprint of particle physics simulations, weather forecast models, and NLP algorithms as well as the carbon effects of distributed computation using multiple CPUs. Finally, we make recommendations on ways for scientists to reduce their GHG emissions as well as discuss the limitations of our approach.

---

[a] **Supplementary Note 1**

# Methods

The carbon footprint of an algorithm depends on two factors: the energy needed to run it and the pollutants emitted when producing such energy. The former depends on the computing resources used (e.g. number of cores, running time, data centre efficiency) while the later, called carbon intensity, depends on the location and production methods used (e.g. nuclear, gas or coal).

There are several competing definitions of "carbon footprint", and in this project, we use the extended definition from Wright et al.[20]. The climate impact of an event is presented in terms of carbon dioxide equivalent ($CO_2e$[b]) and summarises the global warming effect of the GHG emitted in the determined timeframe, here running a set of computations. The GHGs considered are carbon dioxide ($CO_2$), methane ($CH_4$) and nitrous oxide ($N_2O$)[21]; these are the three most common GHGs of the "Kyoto basket" defined in the Kyoto Protocol[22] and represent 97.9% of global GHG emissions[23]. The conversion into $CO_2e$ is done using Global Warming Potential (GWP) factors from the Intergovernmental Panel on Climate Change (IPCC)[21,24] based on a 100-year horizon (GWP100).

When estimating these parameters, accuracy and feasibility must be balanced. This study focuses on a methodology that can be easily and broadly adopted by the community and therefore, restricts the scope of the environmental impact considered to GHGs emitted to power computing facilities for a specific task. Moreover, the framework presented requires no extra computation, nor involves invasive monitoring tools.

## Energy consumption

We model an algorithm's energy[c] needs as a function of the running time, the number, type and process time of computing cores (CPU or GPU), the amount of memory mobilised and the power draw of these resources. The model further includes the efficiency of the data centre[d], i.e. how much extra power is necessary to run the facility (e.g. cooling and lighting).

Similar to previous works[17,18], our estimate is based on the power draw from processors and memory, as well as the efficiency of the data centre. However, we refine the formula and add flexibility by including a unitary power draw (per core and per GB of memory) and the processor's usage factor. We express the energy consumption $E$ (in kilowatt-hours, kWh) as:

$$E = t \times (n_c \times P_c \times u_c + n_m \times P_m) \times PUE \times 0.001 \qquad (1)$$

where $t$ is the running time (hours), $n_c$ the number of cores and $n_m$ the size of memory available (gigabytes). $u_c$ is the core usage factor (between 0 and 1). $P_c$ is the power draw of a computing core and $P_m$ the power draw of the memory (Watt). $PUE$ is the efficiency coefficient of the data centre.

---

[b] Sometimes also called $CO_2eq$, $CO_2equivalent$ or CDE.

[c] Power (in Watt, W) measures the instantaneous draw of a component. Energy (in kilowatt-hours, kWh) measures the power draw over time and is obtained by multiplying the power draw by the running time.

[d] By data centre, we mean the facility hosting the cores and memory, which may not be a dedicated data centre.

The assumptions made regarding the different components are discussed below. It has been previously shown that the power draw of the motherboard is negligible[25].

## Power draw of the computing core

The metric commonly used to report the power draw of a processor, either CPU or GPU, is its thermal design power (TDP, in Watt) and is provided by the manufacturer. TDP values frequently correspond to CPU specifications which include multiple cores, thus here TDP values are normalised to per-core. While TDP is not a direct measure of power consumption, rather the amount of heat a cooling system dissipates during regular use - it is commonly considered a reasonable approximation.

The energy used by the processor is the power draw multiplied by processing time, scaled by the usage factor. However, processing time cannot be known *a priori* and, on some platforms, tracking can be impractical at scale. Modelling exact processing time of past projects may also necessitate re-running jobs, which would generate unnecessary emissions. Therefore, when this processing time is unknown, we make the simplifying assumption that core usage is 100% of run time ($u_c = 1$ in (1)).

## Power draw from memory

Memory power draw is mainly due to background consumption with a negligible contribution from the workload and database size[26]. Moreover, the power draw is mainly affected by the total memory allocated, not by the actual size of the database used, because the load is shared between all memory slots which keeps every slot in a power-hungry active state. Therefore, the primary factor influencing power draw from memory is the quantity of memory mobilised, which simply requires an estimation of the power draw per gigabyte. Measured experimentally, this has been estimated to be 0.3725 W/GB[26,27].

For example, requesting 29GB of memory draws 10.8 W, which is the same as one core of a popular Core-i5 CPU. **Supplementary Figure 1** further compares the power draw of memory to a range of popular CPUs.

## Power draw from storage

The power draw of storage equipment (HDD or SSD) varies significantly with workload[28]. However, in regular use, storage is typically solicited far less than memory and is mainly used as a more permanent record of the data, independently of the task at hand. In idle mode (i.e. not servicing a request but ready to begin the next one), non-optimised HDDs rarely consume more than 6W for 1TB of storage and modern SSDs can draw as little as 0.6W for 800GB of storage[28]. Under conservative assumptions, storage power draw would be 0.006 W/GB. As above, by comparison, the power draw of memory (0.3725 W/GB) and a Core-i5 CPU (10.8W/core) are more than an order of magnitude greater. While the researcher overhead for approximating storage usage may not be substantial, it is unlikely to make a significant difference to overall power usage (and GHG emissions) estimation. Therefore, we do not consider the power consumption of storage in this work.

## Energy efficiency

Data centre energy consumption includes additional factors, such as server cooling systems, power delivery components and lighting. The efficiency of a given data centre can be measured by the Power Usage Effectiveness (PUE)[29,30], defined as the ratio between the total power drawn by the facility and the power used by IT equipment:

$$PUE = \frac{P_{total}}{P_{compute}} \qquad (2)$$

A data centre PUE of 1.0 represents an ideal situation where all power supplied to the building is utilised by computing equipment. The global average of data centres has been estimated as 1.67 in 2019[31]. While data centres with relatively inefficient PUE may not report it as such, some data centres and companies have invested significant resources to bring their PUEs as close to 1.0 as possible; for example, Google has utilised machine learning to reduce its global yearly average PUE to 1.10[32,33].

## Carbon intensity of energy production

For a given country and energy mix, the carbon footprint in $CO_2$e represents the amount of $CO_2$ with the same global warming impact as the GHGs emitted, which simplifies the comparison between different electricity production methods. The carbon footprint of producing 1 kWh of energy is called Carbon Intensity (CI) and varies significantly between locations due to the broad range of production methods (**Supplementary Figure 2**), e.g. from 19 g$CO_2$e/kWh in Switzerland (mainly powered by hydro) to 880 g$CO_2$e/kWh in Australia (mainly powered by coal and gas)[34,35]. We use the 2020 carbon intensity values aggregated by *Carbon Footprint*[35]. These production factors take into account the GHG emissions at the power plants (power generation) as well as, when available, the footprint of distributing energy to the data centre.

## Estimation of carbon footprint

The carbon footprint $C$ (in g$CO_2$e) of producing a quantity of energy $E$ (in kWh) from sources with a carbon intensity $CI$ (in g$CO_2$e/kWh) is then:

$$C = E \times CI \qquad (3)$$

By putting together equations (1) and (3), we obtain the long-form equation of the carbon footprint $C$:

$$C = t \times (n_c \times P_c \times u_c + n_m \times P_m) \times PUE \times CI \times 0.001 \qquad (4)$$

## $CO_2$e of driving and air travel

We contextualise g$CO_2$e by estimating an equivalence in terms of distance travelled by car or by passenger aircraft. Previous studies have estimated the emissions of the average passenger car in Europe as 175 gCO$_2$e/km[21,36] (251 gCO$_2$e/km in the United States[37]). The emissions of flying on a jet aircraft in economy class have been estimated between 139 and 244 gCO$_2$e/km/person, depending on the length of the flight[21]. We use three reference flights: Paris to London (50,000 gCO$_2$e), New York to San Francisco (570,000 gCO$_2$e) and New York to Melbourne (2,310,000 gCO$_2$e)[38].

## CO$_2$ sequestration by trees

Trees play a major role in carbon sequestration and although not all GHGs emitted can be sequestered, CO$_2$ represents 74.4% of these emissions[39]. To provide a metric of reversion for CO$_2$e, we compute the number of trees needed to sequester the equivalent of the emissions of a given computation. We define the metric *tree-months*, the number of months a mature tree needs to absorb a given quantity of CO$_2$. While the amount of CO$_2$ sequestered by a tree per unit of time depends on a number of factors, such as its species, size or environment, it has been estimated that a mature tree sequesters, on average, approximately 11 kg of CO$_2$ per year[40], giving the multiplier in tree-months a value close to 1kg of CO$_2$ per month (0.92g).

## Pragmatic scaling factor

Many analyses are presented as a single run of a particular algorithm or software tool; however, computations are rarely performed only once. Algorithms are run multiple times, sometimes hundreds, systematically or manually, with different parameterisations. Statistical models may include any number of combinations of covariates, fitting procedures, etc. It is important to include these repeats in the carbon footprint. To take into account the number of times a computation is performed in practice, we define the pragmatic scaling factor (PSF), a scaling factor by which the estimated GHG emissions are multiplied.

The value and causes of the PSF vary greatly between tasks. In machine learning, tuning the hyper-parameters of a model requires hundreds, if not thousands[17], of runs, while other tools require less tuning and can sometimes be run a smaller number of times. As per published work or the user's own experience, the PSF should be estimated for any specific task; however, in Green Algorithms we provide for, and recommend that, each user estimate their own PSF.

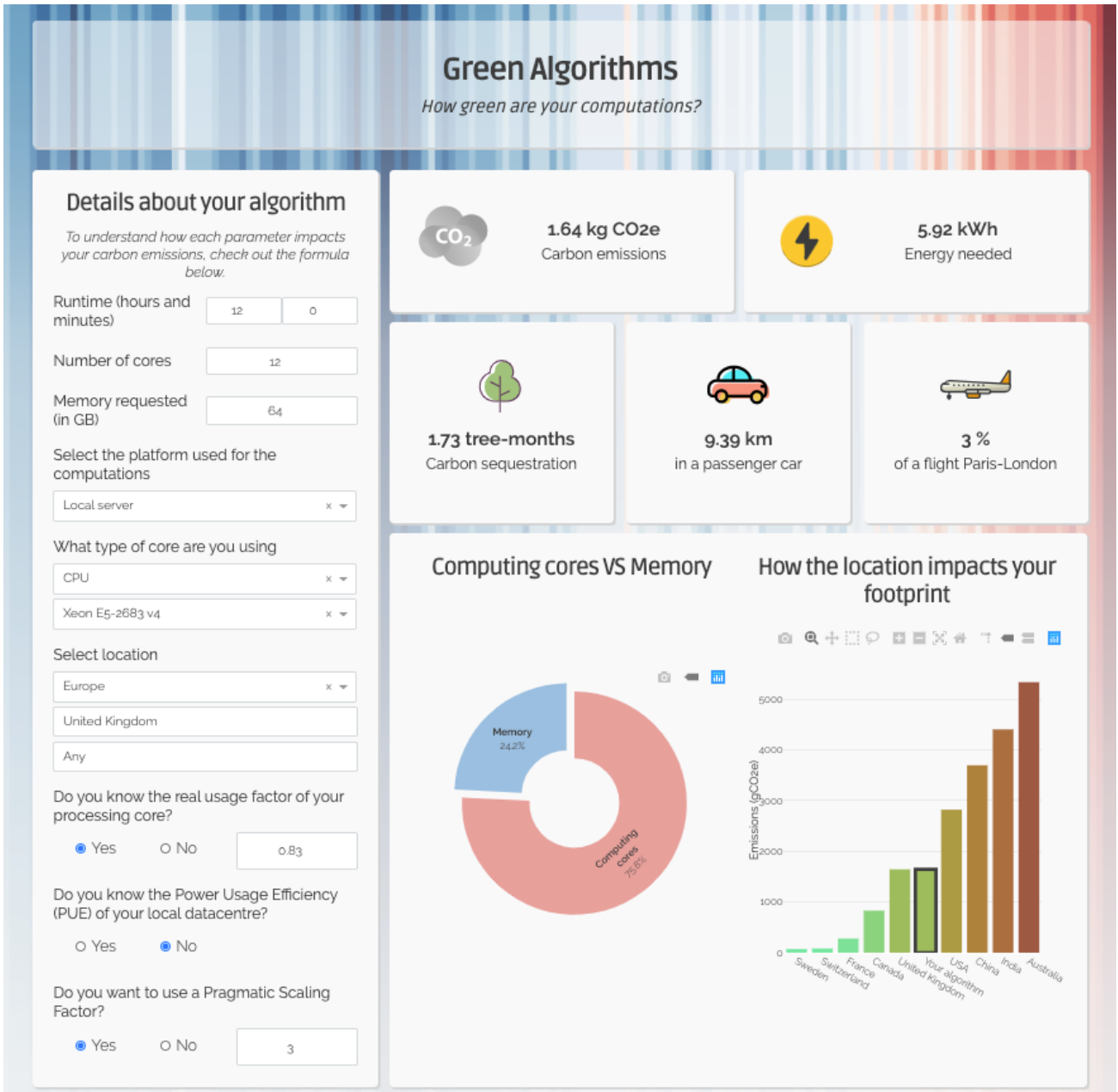

**Figure 1: The Green Algorithms calculator (www.green-algorithms.org)**

# Results

We developed a simple method which estimates the carbon footprint of an algorithm based on a number of factors, including the hardware requirements of the tool, the runtime and the location of the data centre (**Methods**). Using a pragmatic scaling factor, we further augment our model by allowing for empirical estimates of repeated computations for a particular task, e.g. parameter tuning and trial-and-errors. The resultant gCO$_2$e is compared to the amount of carbon sequestered by trees and the emissions of common activities such as driving a car and air travel. We designed a freely available online tool, Green Algorithms (www.green-algorithms.org; **Figure 1**), which implements our approach and allows users to evaluate their computations or estimate the carbon savings or costs of redeploying them on other architectures.

We apply this tool to a range of algorithms selected from a variety of scientific fields: physics (particle simulations and DNA irradiation), atmospheric sciences (weather forecasting), and machine learning (natural language processing) (**Figure 2**). For each task, we curate published benchmarks and use www.green-algorithms.org to estimate the GHG emissions (**Methods**). For parameters independent of the algorithm itself, we use average worldwide values, such as the worldwide average PUE of 1.67[31] and carbon intensity of 475 g$CO_2$e/kWh[41].

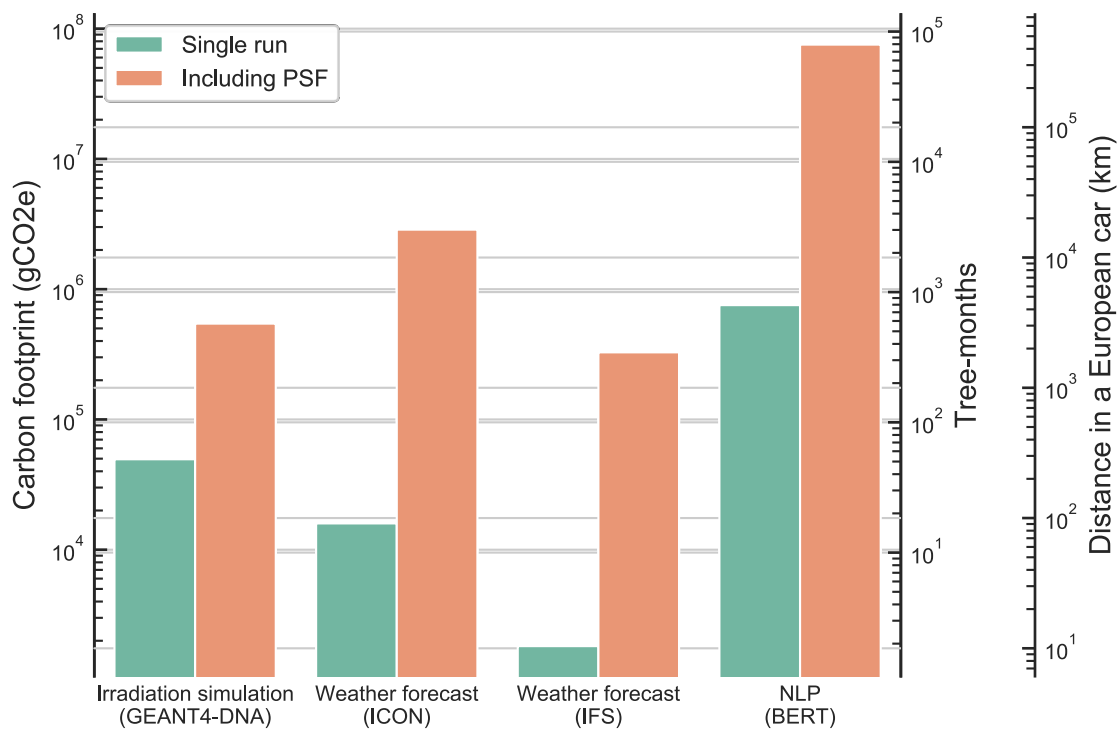

**Figure 2: Carbon footprint (g$CO_2$e) for a selection of algorithms, with and without their Pragmatic Scaling Factor.**

## Particle physics simulations

In particle physics, complex simulations are used to model the passage of particles through matter. Geant4[42] is a popular toolkit based on Monte-Carlo methods with wide-ranging applications, such as the simulation of detectors in the Large Hadron Collider and analysis of radiation burden on patients in clinical practice or external beam therapy[43–45]. Meylan et al.[46] investigated the biological effects of ionising radiations on DNA on an entire human genome ($6.4\times10^9$ nucleotide pairs) using GEANT4-DNA, an extension of GEANT4.

To quantify the DNA damage of radiation, they run experiments with photons of different energy, from 0.5 MeV to 20 MeV. Each experiment runs for three weeks to simulate 5,000 particles (protons) using 24 processing threads and up to 10GB of memory. Using the Green Algorithms tool, and assuming an average CPU power draw (such as the Xeon E5-2680, capable of running 24 threads on 12 cores), and worldwide average values for PUE and carbon intensity, we estimated that a single experiment emits 49,465 g$CO_2$e. When taking into account a PSF of 11, corresponding to the 11 different energy levels tested, the carbon footprint of such study is 544,115 g$CO_2$e. Using estimates of car and air travel (**Methods**), 544,115 g$CO_2$e is approximately equivalent to driving 3,109 km (in a European car) or flying economy from New York to San Francisco. In terms of carbon sequestration (**Methods**), it would take a mature tree 49 years to remove the $CO_2$ equivalent to the GHG emissions of this study from the atmosphere (593 tree-months).

A common way to reduce the running time of algorithms is to distribute the computations over multiple processing cores. If the benefit in terms of time is well documented for each task, as in [47], the environmental impact is usually not taken into account. GEANT4 is a versatile toolbox; it contains an electromagnetic package simulating particle transport in matter and high energy physics detector response[48]. Schweitzer et al.[47] use a standardised example, TestEm12[49], to compare the performances of different hardware configurations, from 1 to 60 cores (i.e. a full Xeon Phi CPU). With the Green Algorithms tool, we estimated the carbon footprint of each configuration (**Figure 3**), which shows that increasing the number of cores up to 15 improves both running time and GHG emissions. However, when multiplying the number of cores further by 4 (from 15 to 60), the running time is only halved, resulting in a two-fold increase in emissions, from 238 to 481 gCO2e. Generally, if the reduction in running time is lower than the relative increase in the number of cores, distributing the computations will worsen the carbon footprint. In particular, scientists should be mindful of marginal improvements in running time which have disproportionally large effects on GHG emissions, as demonstrated by the gap between 30 and 60 cores in **Figure 3**. For any parallelised computation, there is likely to be a specific optimal number of cores for minimal GHG emissions.

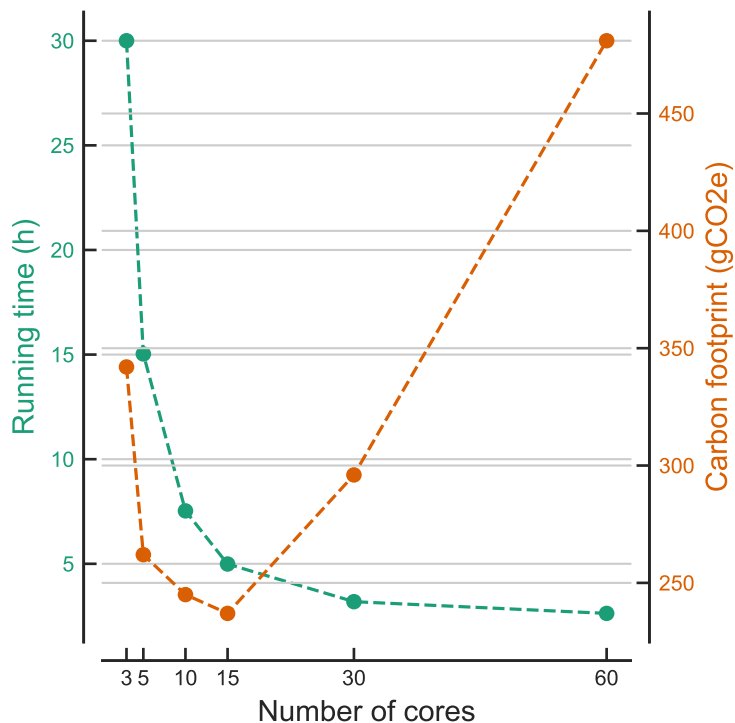

**Figure 3: Effect of parallelisation using multiple cores on run time and carbon footprint using TestEm12 GEANT4 simulation.**

## Weather forecasting

Weather forecasts are based on sophisticated models simulating the dynamics between different components of the earth (such as the atmosphere and oceans). Operational models face stringent time requirements to provide live predictions to the public, with a goal of running about 200-300 forecast days (FDs) in one (wall clock) day[50]. Neumann et al.[50] present the performances of two models in use for current weather forecasts: (i) the Integrated Forecast System (IFS)[51] used by the European Centre for Medium-Range Weather Forecasts (ECMWF) for 10-day forecasts, and (ii) the ICOsahedral Non-hydrostatic (ICON)[52] designed by the German Weather Service (Deutscher Wetterdienst, DWD) and whose predictions are used by more than 30 national weather services[53].

The configurations in daily use by the ECMWF include a supercomputer based in Reading, UK, which has a PUE of 1.45[54], while ICON is run on the German Meteorological Computation Centre (DMRZ)[55] based in Germany (PUE unknown). Neumann et al.[50] ran their experiments on hardware similar to that equipped by both facilities, "Broadwell" CPU nodes (Intel E5-2695v4, 36 cores) and minimum 64GB memory per node. We utilise these parameters for our $CO_2$e emission estimates. It is important to note that ICON and IFS each solve slightly different problems, and therefore are not directly comparable.

The DWD uses ICON with a horizontal resolution of 13km[e] and generates a forecast day in 8 minutes. Based on the experiments run by Neumann et al.[50], this requires 575 Broadwell nodes (20,700 CPU cores). We estimate that generating one forecast day emits 12,848 g$CO_2$e (14 tree-months). With a running time of 8min/FD, ICON can generate 180 forecast days in 24 hours. When taking into account this pragmatic scaling factor of 180, we estimated that each day, the ICON weather forecasting algorithm releases approximately 2,312,653 g$CO_2$e, equivalent to driving 13,215 km or flying from New York to San Francisco four times. In terms of carbon sequestration, the emissions of each day of ICON weather forecast are equivalent to 2,523 tree-months.

At ECMWF, IFS makes 10-day operational weather forecasts with a resolution of 9km. To achieve a similar threshold of 180 FDs/day, 128 Broadwell nodes are necessary (4,608 cores)[50,56]. Using the PUE of the UK ECMWF facility (1.45), we estimate the impact of producing one forecast day with IFS to be 1,660 g$CO_2$e. Using a PSF of 180 for one day's forecasts, we estimated emissions of 298,915 g$CO_2$e, equivalent to driving 1,708 km or three return flights between Paris and London. These emissions are equivalent to 326 tree-months.

Furthermore, we modelled the planned scenario of the ECMWF transferring its supercomputing to Bologna, Italy, in 2021[57]. Compared to the data centre in Reading, the new data centre in Bologna is estimated to have a more efficient PUE of 1.27[58]. *Prima facie* this move appears to save substantial GHG emissions; however, it is notable that the carbon intensity of Italy is 33% higher than the UK[35]. Unless the sources of electricity for the data centre in Bologna are different from the rest of Italy and in the absence of further optimisations, we estimated that the move would result in an 18% increase in GHG emissions from the ECMWF (from 298,915 to 350,063 g$CO_2$e).

## Natural language processing

In natural language processing (NLP), the complexity and financial costs of model training are major issues[16]. This has motivated the development of language representations that can be trained once to model the complexity of natural language, and which could be used as input for more specialised algorithms. The BERT (Bidirectional Encoder Representations from Transformers)[59] algorithm is a field leader which yields both high performance and flexibility: state-of-the-art algorithms for more specific tasks are obtained by fine-tuning a pre-trained BERT model, for example in scientific text analysis[60] or biomedical text mining[61]. Yet, while the BERT model is intended to avoid retraining, many data scientists, perhaps understandably, continue to recreate or attempt to improve upon BERT, leading to redundant and ultimately inefficient computation as well as excess $CO_2e$ emissions. Even with optimised hardware (such as NVIDIA Volta GPUs), a BERT training run may take three days or more[62].

---

[e] The horizontal resolution represents the level of geographical detail achieved when modelling the different weather phenomenon.

Using these optimised parameters, Strubell et al.[17] showed that a run time of 79 hours on 64 Tesla V100 GPUs was necessary to train BERT, with a usage factor of the GPUs of 62.7%. With the Greens Algorithms calculator, we estimated that a BERT training run would emit 754,407 $gCO_2e$ (driving 4,311 km in a European car; 1.3 flights from New York to San Francisco; and 823 tree-months). When considering a conservative PSF of 100 for hyperparameters search, we obtain a carbon footprint of 75,440,740 $gCO_2e$.

While BERT is a particularly widely utilised NLP tool, Google has also developed a chatbot algorithm, Meena, which was trained for 30 days on a TPU-v3 Pod containing 2,048 Tensor Processing Unit (TPU) cores[63]. There is limited information on the power draw of TPU cores and memory; however, the power supply of this pod has been estimated to be 288 kW[64]. Using a run time of 30 days, assuming full usage of the TPUs and ignoring memory power draw, the Greens Algorithms calculator estimated that Meena training emitted 164,488,320 $gCO_2e$, which corresponds to 179,442 tree-months or 71 flights between New-York and Melbourne.

## Discussion

The method and Green Algorithms tool presented here provides users with a practical way to estimate the carbon footprint of their computations. The method focuses on producing sensible estimates with small overheads for scientists wishing to measure the footprint of their work. Consequently, the online calculator is simple to use and generalisable to nearly any computational task. We applied the Green Algorithms calculator to a variety of tasks, including particle physics simulations, weather forecasting and natural language processing, to estimate their relative and ongoing carbon emissions. Real-world changes to computational infrastructures, such as moving data centres, was also quantifiable in terms of carbon footprint and was shown to be of substantive importance, e.g. moving data centres may attain a more efficient PUE but a difference in carbon intensity may negate any efficiency gains, potentially making such a move detrimental to the environment.

Our work substantially enhances and extends prior frameworks for estimating the carbon footprint of computation. In particular, we have integrated and formalised previously unclear factors such as usage factor and unitary power draw (per-core or per-GB of memory). As a result, and as presented in the **Methods**, the carbon footprint of an algorithm can be broken down to a small number of key, easily quantifiable elements, such as number of cores, memory size and usage factor. This reduces the burden on the user, who is not required to either measure the power draw of hardware manually or use a limited range of cloud providers for their computations. This makes the method highly flexible in comparison to previous work. Besides drawing attention to the growing issue of GHG emissions of data centres, one of the benefits of presenting a detailed open methodology and tool is to provide users with the information they need to reduce their carbon footprint. Perhaps the most important challenge in green computing is to make the estimation and reporting of GHG emissions a standard practice. This requires transparent and easy-to-use methodology, such as the Green Algorithms calculator (www.green-algorithms.org) and open-source code and data presented here (see **Code availability**).

Our approach has a number of limitations. First, the carbon footprint estimated is restricted to GHGs emitted to power computers during a particular task. We do not perform a Life Cycle Assessment (LCA) and therefore, do not consider the full environmental and social impact of manufacturing, maintaining and disposing of the hardware used, or the maintenance of the power plants. Including these is impractical at scale and would greatly reduce who can use the method. Besides, the conversion of the impact of various GHG into $CO_2e$ is commonly based on a 100-year timescale; however, this is now debated as it can misrepresent the impact of short-lived climate pollutants like methane[65] and new standards may be needed in the future. Second, the TDP may substantially underestimate power draw in some situations. For example, when hyperthreading is used, the real power consumption can be double the indicated TDP[66]. The TDP value remains a sensible estimate of the base consumption of the processor in most situations, but users using hyperthreading should be aware of the impact on power consumption. Third, while the power consumption from storage is usually minimal at the scale of one computation, if central storage is constantly queried by the algorithm (e.g. to avoid overloading memory), this can be an important factor in power draw; however, there are resources which can be utilised if the algorithm is designed to be heavily storage reliant[28]. Moreover, at the scale of the data centre, storage represents a significant part of electricity usage[28] and research projects relying on large databases should separately acknowledge the long-term carbon footprint of storage. Fourth, while some averaging is necessary, the energy mix of a country varies by the hour. For example, the carbon intensity of South Australia, which relies on wind and gas to produce electricity[67], can vary between 112 and 592 $gCO_2e$/kWh within one day, depending on the quantity of coal-produced electricity imported from the neighbouring state of Victoria[34]. Although most regions are relatively stable, these outliers may require a finer estimation. Our online calculator uses averaged values sourced from government reports[35]. Fifth, the PUE has some limitations as a measure of data centres energy usage[68,69], due to inconsistencies in ways to calculate it. For example, reporting of PUE is highly variable from yearly averages to best-case scenarios, e.g. in winter when minimal cooling is required (as demonstrated by Google's quarterly results[32]). Whether to include infrastructure components such as security or on-site power generation is also source of discrepancies between data centres[30]. Although some companies present well-justified results, many PUEs have no or insufficient justification. Furthermore, PUE is not defined when computations are run on a laptop or desktop computer. As the device is used for multiple tasks simultaneously, it is impossible to estimate the power overhead due to the algorithm. In the calculator, we use a PUE of 1 because of the lack of information, but we caution this should not be interpreted as a sign of efficiency. Even though discrepancies will remain, the widespread adoption of an accurate, transparent and certified estimation of PUE, such as the ISO/IEC standard[70], would be a substantial step for the computing community. Sixth, the carbon emissions in the **Results** are based on manual curation of the literature. When parameters such as usage factor or PUE were not specified, we made some assumptions (100% core usage, or using average PUE) that can explain differences between our estimates and the real emissions. For best results, authors should estimate and publish their emissions.

There are various, realistic actions one can take to reduce the carbon footprint of their computation. Acting on the various parameters in Green Algorithms (see **Methods**), is a clear and easy way approach. Below, we describe a selection of practical changes one can make:

**Algorithm optimisation:** Increasing the efficiency of an algorithm can have myriad benefits, even apart from reducing its carbon footprint. Therefore, we highly recommend this and foresee algorithm optimisation as one of the most productive, easily recognisable core activities of green computing. While speed is an obvious efficiency gain, part of algorithm optimisation also includes memory minimisation. The power draw from memory mainly depends on the memory requested, not the actual memory used[26], and the memory requested is often the peak memory needed for one step of the algorithm (typically a merge or aggregation). By optimising these steps, one can easily reduce energy consumption.

**Reduce the Pragmatic Scaling Factor:** Limiting the number of times an algorithm runs, especially those that are power hungry, is perhaps the easiest way to reduce carbon footprint. Relatedly, best practices to limit PSF (as well as financial cost) include limiting parameter fine-tuning to the minimum necessary and building a small-scale example for debugging.

**Choice of data centre:** Carbon footprint is directly proportional to data centre efficiency and the carbon intensity of the location. The latter is perhaps the parameter which most affects total carbon footprint because of inter-country variation, from under 20 gCO2e/kWh in Norway and Switzerland to over 800 gCO2e/kWh in Australia, South Africa and some US states. To rigorously assess the impact of punctually relocating computations, the marginal carbon intensity, rather than the average one, should be used[71]. The marginal value depends on which power plant would be solicited to meet the unexpected increased demand. Although it would ideally be used, it varies by the hour and is often not practical to estimate accurately at scale. When the marginal carbon intensity is unknown, the average one (presented in **Methods** and **Supplementary Figure 2**) can be used by scientists as a practical lower bound estimate to assess the benefit of moving computations. Indeed, due to the low operating cost of renewable technologies, the marginal power plants (which are the last one solicited) are generally high-carbon technologies such as fuel or gas[71] which leads the marginal CI to be higher than the average CI. Besides, if the move is permanent, by relocating an HPC facility or using cloud computing for example, then the energy needs are incorporated into utility planning and the average carbon intensity is the appropriate metric to use. Data centre efficiency (PUE) varies widely between facilities but, in general, large data centres optimise cooling and power supply, reducing the energy overhead and make them more efficient than personal servers. Notably, a 2016 report estimated that if 80% of small US data centres were aggregated into hyperscale facilities, energy usage would reduce by 25%[72]. For users to make informed choices, data centres should report their PUE and other energy metrics. While large providers like Google or Microsoft widely advertise their servers' efficiency[32,73], smaller structures often do not.

**Offsetting GHG emissions:** Carbon offsetting is a flexible way to compensate for carbon footprint. An institution or a user themself can directly support reductions in $CO_2$ or other greenhouse gases, e.g. fuel-efficient stoves in developing countries, reducing deforestation or hydroelectric or wind-based power plants[74,75]. The pros and cons of carbon offsetting are still debated due to the variety of mechanisms and intricate international legislations and competing standards. Therefore, we only present here an overview and point interested scientists at some resources. Multiple international standards regulate the purchase of carbon credits and ensure the efficiency of the projects supported[76]. Most of the well-established standards are managed by non-profits and abide by the mechanisms set in place by the Kyoto protocol (in particular Certified Emission Reduction)[77] and the PAS 2060 Carbon Neutrality standard from the British Standards Institution[78]. Although the primary aim is carbon offsetting, projects are often also selected in line with the United Nations' Agenda 30 for Sustainable Development[79], a broader action plan addressing inequalities, food security and peace. Amongst the most popular standards are the Gold Standard (founded by WWF and other NGOs)[80], Verra (formerly Verified Carbon Standard)[81] and the American Carbon Registry (a private voluntary greenhouse gas registry)[82]. In addition to direct engagement with these standards, platforms like Carbon Footprint[74] select certified projects and facilitate the purchase of credits.

## Conclusions

The framework presented here is generalisable to nearly any computation and may be used as a foundation for other aspects of green computing. The carbon footprint of computation is substantial and may be affecting the climate. We therefore hope that this new tool and metrics raise awareness of these issues as well as facilitate pragmatic solutions which may help to mitigate the environmental consequences of modern computation. Overall, with the right tools and practices, we believe HPC and cloud computing can be immensely positive forces for both improving the human condition and saving the environment.

## Data availability

All data used for the calculator is available on GitHub: https://github.com/GreenAlgorithms/green-algorithms-tool/tree/master/data.

## Code availability

All code supporting the calculator is available on GitHub: https://github.com/GreenAlgorithms/green-algorithms-tool

14. Gai, K., Qiu, M., Zhao, H., Tao, L. & Zong, Z. Dynamic energy-aware cloudlet-based mobile cloud computing model for green computing. *J. Netw. Comput. Appl.* **59**, 46–54 (2016).

15. Goodfellow, I., Bengio, Y. & Courville, A. *Deep Learning*. (MIT Press, 2016).

16. Schwartz, R., Dodge, J., Smith, N. A. & Etzioni, O. Green AI. *ArXiv190710597 Cs Stat* (2019).

17. Strubell, E., Ganesh, A. & McCallum, A. Energy and Policy Considerations for Deep Learning in NLP. *ArXiv190602243 Cs* (2019).

18. Lacoste, A., Luccioni, A., Schmidt, V. & Dandres, T. Quantifying the Carbon Emissions of Machine Learning. *ArXiv191009700 Cs* (2019).

19. Henderson, P. *et al.* Towards the Systematic Reporting of the Energy and Carbon Footprints of Machine Learning. *ArXiv200205651 Cs* (2020).

20. Wright, L. A., Kemp, S. & Williams, I. 'Carbon footprinting': towards a universally accepted definition. *Carbon Manag.* **2**, 61–72 (2011).

21. Hill, N. *et al.* 2020 Government greenhouse gas conversion factors for company reporting: Methodology paper. *Dep. Bus. Energy Ind. Strategy* 128 (2020).

22. *Kyoto Protocol To The United Nations Framework Convention On Climate Change*. (1997).

23. Ritchie, H. & Roser, M. $CO_2$ and Greenhouse Gas Emissions. *Our World Data* (2020).

24. Fourth Assessment Report — IPCC. https://www.ipcc.ch/assessment-report/ar4/.

25. Geng, H. *Data Center Handbook*. (John Wiley & Sons, 2014).

26. Karyakin, A. & Salem, K. An analysis of memory power consumption in database systems. in *Proceedings of the 13th International Workshop on Data Management on New Hardware - DAMON '17* 1–9 (ACM Press, 2017). doi:10.1145/3076113.3076117.

27. Angelini, C., August 29, I. W. & 2014. Intel Core i7-5960X, -5930K And -5820K CPU Review: Haswell-E Rises. *Tom's Hardware* https://www.tomshardware.com/uk/reviews/intel-core-i7-5960x-haswell-e-cpu,3918-13.html.

28. Tomes, E. & Altiparmak, N. A Comparative Study of HDD and SSD RAIDs' Impact on Server Energy Consumption. in *2017 IEEE International Conference on Cluster Computing (CLUSTER)* 625–626 (2017). doi:10.1109/CLUSTER.2017.103.

29. Belady, C. L. & Malone, C. G. Metrics and an Infrastructure Model to Evaluate Data Center Efficiency. in 751–755 (American Society of Mechanical Engineers Digital Collection, 2010). doi:10.1115/IPACK2007-33338.

30. Avelar, V., Azevedo, D., French, A. & Power, E. N. PUE: a comprehensive examination of the metric. *White Pap.* **49**, (2012).

31. Andy Lawrence. Is PUE actually going UP? *Uptime Institute Blog* https://journal.uptimeinstitute.com/is-pue-actually-going-up/ (2019).

32. Efficiency – Data Centers – Google. *Google Data Centers* https://www.google.com/about/datacenters/efficiency/.

33. Gao, J. *Machine Learning Applications for Data Center Optimization*. (2014).

34. Tranberg, B. *et al.* Real-time carbon accounting method for the European electricity markets. *Energy Strategy Rev.* **26**, 100367 (2019).

35. carbonfootprint.com - International Electricity Factors. https://www.carbonfootprint.com/international_electricity_factors.html.
16

# Acknowledgements

LL was supported by the University of Cambridge MRC DTP (MR/S502443/1). JG was supported by a La Trobe University Postgraduate Research Scholarship jointly funded by the Baker Heart and Diabetes Institute and a La Trobe University Full-Fee Research Scholarship. This work was supported by core funding from: the UK Medical Research Council (MR/L003120/1), the British Heart Foundation (RG/13/13/30194; RG/18/13/33946) and the National Institute for Health Research [Cambridge Biomedical Research Centre at the Cambridge University Hospitals NHS Foundation Trust] [*]. This work was also supported by Health Data Research UK, which is funded by the UK Medical Research Council, Engineering and Physical Sciences Research Council, Economic and Social Research Council, Department of Health and Social Care (England), Chief Scientist Office of the Scottish Government Health and Social Care Directorates, Health and Social Care Research and Development Division (Welsh Government), Public Health Agency (Northern Ireland), British Heart Foundation and Wellcome. MI was supported by the Munz Chair of Cardiovascular Prediction and Prevention. This study was supported by the Victorian Government's Operational Infrastructure Support (OIS) program.

*The views expressed are those of the authors and not necessarily those of the NHS, the NIHR or the Department of Health and Social Care.

# Supplementary Materials

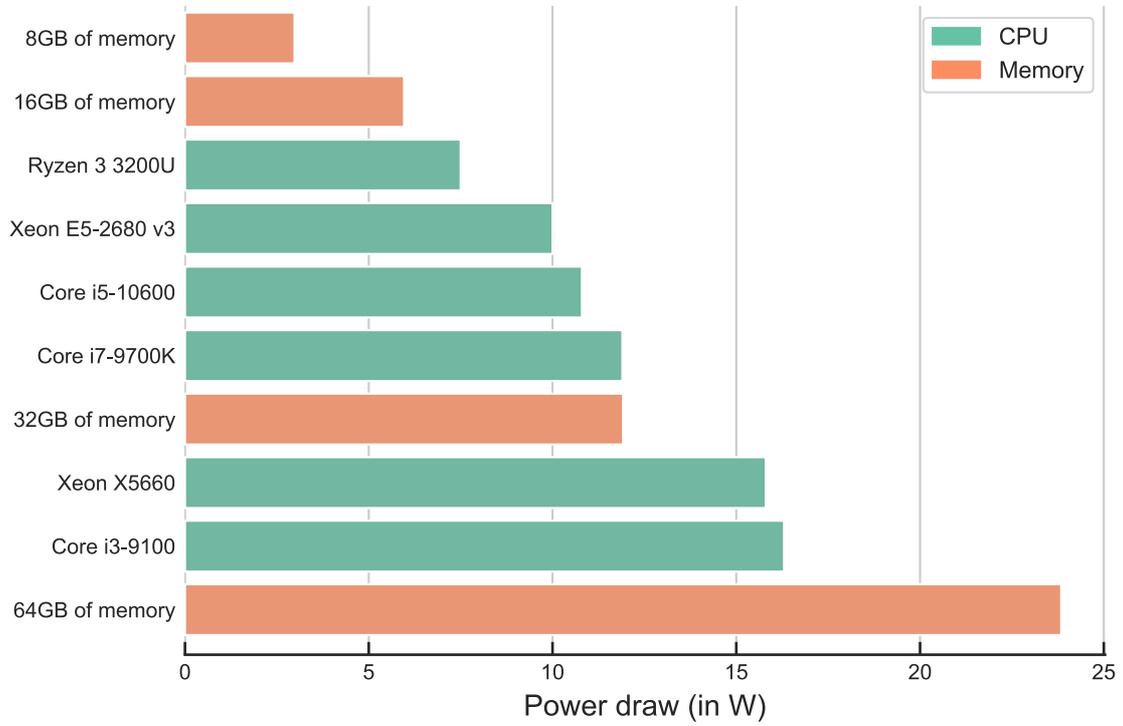

**Supplementary Figure 1: Comparison of power draw (per core) between popular CPUs and memory.**

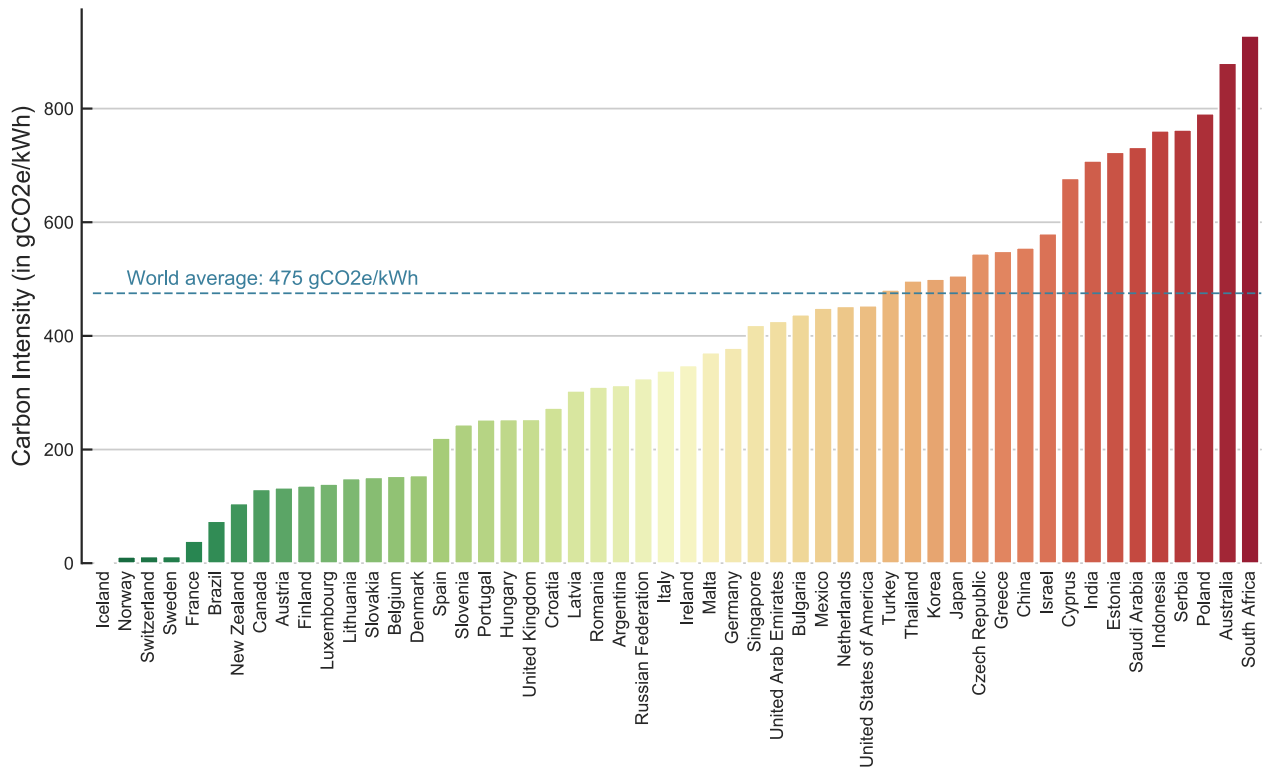

**Supplementary Figure 2: Worldwide carbon intensity distribution by countries, curated from Carbon Footprint[35].**

## Supplementary Note 1

We estimated the annual GHG emissions of data centres to be around 100 Mt $CO_2$e using two distinct approaches. First, using the total electricity demand of data centres, estimated to be around 200 TWh[1] (1% of global demand) and the world average carbon intensity (475 g$CO_2$e/kWh[41]). Equation (3) (**Methods**) gives a total footprint of $95 \times 10^6$ t$CO_2$e. Another estimation is based on data centres being responsible for 0.3% of global emissions[1] (0.3% of $36 \times 10^9$ t$CO_2$e[23]), which yields $108 \times 10^6$ t$CO_2$e.